\documentclass[conference]{IEEEtran}
\usepackage{subcaption}
\usepackage{xcolor}
\usepackage{ifthen}
\usepackage{tikz}
\usepackage{pgfplots,pgfplotstable}
\pgfplotsset{
        set layers={
            background,
            main,
        },
    }
\newlength{\figheight}
\usepackage[square,sort,comma,numbers]{natbib}
\usepackage[T1]{fontenc}

\usetikzlibrary{calc, patterns, spy}

\usepackage{caption,subcaption}
\usepackage[binary-units=true]{siunitx} 
\usepackage[frozencache=true,cachedir=minted-main]{minted}
\usepackage{listings}
\setminted{fontsize=\small,baselinestretch=1}
\usepackage{booktabs}
\usepackage[colorlinks=true,linkcolor=black,anchorcolor=black,citecolor=black,filecolor=black,menucolor=black,runcolor=black,urlcolor=black]{hyperref}
\setcitestyle{square}

\usepackage[ruled, lined, linesnumbered, commentsnumbered, longend]{algorithm2e}

\definecolor{airforceblue}{rgb}{0.36, 0.54, 0.66}

\definecolor{pairedOneLightBlue}{HTML}{a6cee3}
\definecolor{pairedTwoDarkBlue}{HTML}{1f78b4}
\definecolor{pairedThreeLightGreen}{HTML}{b2df8a}
\definecolor{pairedFourDarkGreen}{HTML}{33a02c}
\definecolor{pairedFiveLightRed}{HTML}{fb9a99}
\definecolor{pairedSixDarkRed}{HTML}{e31a1c}
\definecolor{butter1}{rgb}{0.988,0.914,0.310}
\definecolor{butter2}{rgb}{0.929,0.831,0.000}
\definecolor{butter3}{rgb}{0.769,0.627,0.000}
\definecolor{orange1}{rgb}{0.988,0.686,0.243}
\definecolor{orange2}{rgb}{0.961,0.475,0.000}
\definecolor{orange3}{rgb}{0.808,0.361,0.000}
\definecolor{chocolate1}{rgb}{0.914,0.725,0.431}
\definecolor{chocolate2}{rgb}{0.757,0.490,0.067}
\definecolor{chocolate3}{rgb}{0.561,0.349,0.008}
\definecolor{chameleon1}{rgb}{0.541,0.886,0.204}
\definecolor{chameleon2}{rgb}{0.451,0.824,0.086}
\definecolor{chameleon3}{rgb}{0.306,0.604,0.024}
\definecolor{skyblue1}{rgb}{0.447,0.624,0.812}
\definecolor{skyblue2}{rgb}{0.204,0.396,0.643}
\definecolor{skyblue3}{rgb}{0.125,0.290,0.529}
\definecolor{plum1}{rgb}{0.678,0.498,0.659}
\definecolor{plum2}{rgb}{0.459,0.314,0.482}
\definecolor{plum3}{rgb}{0.361,0.208,0.400}
\definecolor{scarletred1}{rgb}{0.937,0.161,0.161}
\definecolor{scarletred2}{rgb}{0.800,0.000,0.000}
\definecolor{scarletred3}{rgb}{0.643,0.000,0.000}
\definecolor{aluminium1}{rgb}{0.933,0.933,0.925}
\definecolor{aluminium2}{rgb}{0.827,0.843,0.812}
\definecolor{aluminium3}{rgb}{0.729,0.741,0.714}
\definecolor{aluminium4}{rgb}{0.533,0.541,0.522}
\definecolor{aluminium5}{rgb}{0.333,0.341,0.325}
\definecolor{aluminium6}{rgb}{0.180,0.204,0.212}

\definecolor{blind_safe_one_scheme_three_colors}{RGB}{102,194,165}
\definecolor{blind_safe_two_scheme_three_colors}{RGB}{252,141,98}
\definecolor{blind_safe_three_scheme_three_colors}{RGB}{141,160,203}

\definecolor{blind_safe_one_scheme_four_colors}{RGB}{166,206,227}
\definecolor{blind_safe_two_scheme_four_colors}{RGB}{31,120,180}
\definecolor{blind_safe_three_scheme_four_colors}{RGB}{178,223,138}
\definecolor{blind_safe_four_scheme_four_colors}{RGB}{51,160,44}

\definecolor{blind_safe_one_scheme_five_colors}{RGB}{240,249,232}
\definecolor{blind_safe_two_scheme_five_colors}{RGB}{186,228,188}
\definecolor{blind_safe_three_scheme_five_colors}{RGB}{123,204,196}
\definecolor{blind_safe_four_scheme_five_colors}{RGB}{67,162,202}
\definecolor{blind_safe_five_scheme_five_colors}{RGB}{8,104,172} 

\definecolor{blind_safe_one_scheme_seven_colors_grnblu}{RGB}{240,249,232}
\definecolor{blind_safe_two_scheme_seven_colors_grnblu}{RGB}{204,235,197}
\definecolor{blind_safe_three_scheme_seven_colors_grnblu}{RGB}{168,221,181}
\definecolor{blind_safe_four_scheme_seven_colors_grnblu}{RGB}{123,204,196}
\definecolor{blind_safe_five_scheme_seven_colors_grnblu}{RGB}{78,179,211}
\definecolor{blind_safe_six_scheme_seven_colors_grnblu}{RGB}{43,140,190}
\definecolor{blind_safe_seven_scheme_seven_colors_grnblu}{RGB}{8,88,158}

\definecolor{blind_safe_one_scheme_seven_colors}{RGB}{118,42,131}
\definecolor{blind_safe_two_scheme_seven_colors}{RGB}{175,141,195}
\definecolor{blind_safe_three_scheme_seven_colors}{RGB}{231,212,232}
\definecolor{blind_safe_four_scheme_seven_colors}{RGB}{247,247,247}
\definecolor{blind_safe_five_scheme_seven_colors}{RGB}{217,240,211}
\definecolor{blind_safe_six_scheme_seven_colors}{RGB}{127,191,123}
\definecolor{blind_safe_seven_scheme_seven_colors}{RGB}{27,120,55}

\definecolor{yellow_one}{RGB}{255,255,212}
\definecolor{yellow_two}{RGB}{254,217,142}
\definecolor{yellow_three}{RGB}{254,153,41}
\definecolor{yellow_four}{RGB}{217,95,14}
\definecolor{yellow_five}{RGB}{153,52,4}

\definecolor{highlight_code_color1}{RGB}{199,233,180}
\definecolor{highlight_code_color2}{RGB}{254,196,79}

\makeatletter
\lstdefinelanguage{mlir}{
    classoffset=0,
    morekeywords={
        module,
        func,
        cinm,
        cnm,
        cim
    },
    morestring=[b]",
    alsoletter={\%},
    keywordsprefix={\%}
}

\makeatletter
\newcommand{\linebreakand}{%
  \end{@IEEEauthorhalign}
  \hfill\mbox{}\par
  \mbox{}\hfill\begin{@IEEEauthorhalign}
}
\makeatother

\lstdefinelanguage{Python}{
  keywords={and,as,assert,break,class,continue,def,del,elif,else,%
    except,exec,finally,for,from,global,if,import,in,is,lambda,%
    not,or,pass,print,raise,return,try,while,with,yield, False, self},
  keywordstyle=\color{blue}\bfseries,
  sensitive=true,
  commentstyle=\color{gray},
  basicstyle=\ttfamily,
  morestring=[b]',
  morestring=[b]",
  morecomment=[l]\#,
  morecomment=[s]{'''}{'''},
  morecomment=[s]{"""}{"""},
  emph={return}, 
  emphstyle=\color{magenta},
}

\newcommand{\hlhlhl}{\makebox[0pt][l]{\color{highlight_code_color1}\rule[-2pt]{0.93\linewidth}{7pt}}}
\newcommand{\glglgl}{\makebox[0pt][l]{\color{highlight_code_color2}\rule[-2pt]{0.93\linewidth}{7pt}}}

\newcommand{\hlhlhlmedium}{\makebox[0pt][l]{\color{highlight_code_color1}\rule[-2pt]{0.9\linewidth}{7pt}}}

\newcommand{\hlhlhlsmall}{\makebox[0pt][l]{\color{highlight_code_color1}\rule[-2pt]{0.80\linewidth}{7pt}}}
\newcommand{\glglglsmall}{\makebox[0pt][l]{\color{highlight_code_color2}\rule[-2pt]{0.80\linewidth}{7pt}}}

\begin{document}

\title{C4CAM: A Compiler for CAM-based In-memory Accelerators}

\author{\IEEEauthorblockN{Hamid Farzaneh}
\IEEEauthorblockA{TU Dresden, Germany\\
hamid.farzaneh@tu-dresden.de
}
\and 
\IEEEauthorblockN{João Paulo C. de Lima}
\IEEEauthorblockA{TU Dresden/ScaDS.AI, Germany\\
joao.lima@tu-dresden.de
}
\and 
\IEEEauthorblockN{Mengyuan Li}
\IEEEauthorblockA{University of Notre Dame, USA\\
mli22@nd.edu 
}
\linebreakand
\IEEEauthorblockN{Asif Ali Khan}
\IEEEauthorblockA{TU Dresden, Germany\\
asif\_ali.khan@tu-dresden.de
}
\and
\IEEEauthorblockN{Xiaobo Sharon Hu}
\IEEEauthorblockA{University of Notre Dame, USA\\
shu@nd.edu 
}
\and 
\IEEEauthorblockN{Jeronimo Castrillon}
\IEEEauthorblockA{TU Dresden/ScaDS.AI, Germany\\
jeronimo.castrillon@tu-dresden.de
}
}


\maketitle

\begin{abstract}
Machine learning and data analytics applications increasingly suffer from the high latency and energy consumption of conventional von Neumann architectures.
Recently, several in-memory and near-memory systems have been proposed to remove this \emph{von Neumann} bottleneck.
Platforms based on \emph{content-addressable memories} (CAMs) are particularly interesting due to their efficient support for the search-based operations that form the foundation for many applications, including K-nearest neighbors (KNN), high-dimensional computing (HDC), recommender systems, and one-shot learning among others.
Today, these platforms are designed by hand and can only be programmed with low-level code, accessible only to hardware experts.  
In this paper, we introduce C4CAM, the first compiler framework to quickly explore CAM configurations and to seamlessly generate code from high-level TorchScript code. 
C4CAM employs a hierarchy of abstractions that progressively lowers programs, allowing code transformations at the most suitable abstraction level. 
Depending on the type and technology, CAM arrays exhibit varying latencies and power profiles. 
Our framework allows analyzing the impact of such differences in terms of system-level performance and energy consumption, and thus supports designers in selecting appropriate designs for a given application.  
\end{abstract}


\section{Introduction}
\label{sec:intro}

Search operations come in numerous forms at the heart of many comparison-intensive applications. In the past decade, the revolution in machine learning, data analytics, and bioinformatics  has played a significant role in driving the demand for efficient hardware acceleration of these operations.
Domains such as network security \cite{dlugosch2014efficient}, bioinformatics \cite{roy2015discovering}, data mining and data analytics \cite{yu2017robotomata} heavily rely on \emph{exact matching} of the query pattern with pre-stored patterns. 
In other applications, such as K-nearest neighbors (KNN) and genome analysis~\cite{ni2019ferroelectric,hanhan2022edam}, the emphasis lies on identifying similarities rather than exact pattern matching. 
In approximate search, when the dissimilarity between a stored pattern and the query pattern is within a predefined threshold, the stored pattern is regarded as a ``match``.
From the computational standpoint, both exact and approximate search operations are time-consuming and are often bottlenecks in comparison-intensive kernels~\cite{yin2021deep}.

Recently, there has been a surge in the adoption of \emph{content addressable memories} {CAM}-based system designs for efficient search operations.
CAMs were originally used in network routing and CPU caching \cite{pagiamtzis2006content}. Recently, they found applications in a wider range of data-intensive domains~\cite{li2022imars,ni2019ferroelectric,hanhan2022edam}.
CAMs allow massively parallel search operations for an input query, enabling the search to be performed across the entire memory with a single operation.
CAM's high-speed parallel search makes it a popular component for constructing cutting-edge \emph{compute-in-memory} (CIM) systems, aiming to provide an energy-efficient alternative to the von Neumann bottleneck in terms of both latency and energy consumption.

CAM designs are broadly classified into binary, ternary, multi-state, and analog CAMs (BCAM, TCAM, MCAM, ACAM, respectively), with implementations based on either conventional CMOS or emerging non-volatile memory (NVM) technologies ~\cite{liu2023reconfigurable,yin2021deep,ali2020ramann,ni2019ferroelectric}.
Compared to CMOS technologies, NVM technologies, like magnetic RAM (MRAM), resistive RAM (ReRAM) or ferroelectric (FeFET), are denser and more energy efficient, yielding more efficient CAM arrays \cite{narla2022modeling,graves2020memory,hu2021memory}.
BCAMs and TCAMs use a bit-wise Hamming distance (HD) to compare the query and stored data, whereas MCAMs and ACAMs apply a specific distance metric to compare the query with memory entries and determine which memory entries match the query based on the distance metric. 
In terms of match types, CAMs can be classified into the exact match (EX), best match (BE), and threshold match (TH)~\cite{hu2021memory}.

Although CAM designs have shown better performance than traditional methods for computing similarity in many domains~\cite{li2022imars,ni2019ferroelectric,hanhan2022edam}, effectively mapping applications written in high-level programming languages onto CAM-based accelerators remains a challenge. 
This is due to the disparity in the abstractions of the applications (high-level) and the (low-level) set of commands needed to program the CAM arrays. 
Presently, CAM arrays are programmed manually with low-level code that only the device experts understand. 
Existing design automation and compilation tools for in-memory computing~\cite{soeken2016mig,siemieniuk2021occ} do not provide support for CAM primitives, highlighting the need for solutions that can support mapping a wider range of applications and accelerate the design process.

This paper proposes C4CAM, the first end-to-end automated framework that enables efficient mapping of applications from a higher TorchScript program onto CAM arrays. C4CAM leverages the multi-level intermediate representation (MLIR) framework to seamlessly optimize and offload comparison-intensive kernels to CAM-enabled systems. 

Concretely, we make the following contributions:

\begin{itemize}
    \item An automated end-to-end compilation flow that (i) makes CAM accelerators accessible to non-experts and 
    (ii) enables device/circuit/architecture experts to explore design trade-offs. 
    C4CAM takes applications written in TorchScript along an architectural model for retargetability and generates code for the given architecture (see Section~\ref{sec:camcompiler}). 
    \item An extension to the MLIR front-end to express search operations in PyTorch applications (see Section~\ref{subsec:c4cam-frontEnd}). 
    \item We extend the CIM abstraction from \cite{khan2023cinm} to cater for CAM accelerators. Specifically, we propose analyses to detect computational primitives in applications that can be rewritten as search operations (see Section~\ref{subsec:cimdialect}). 
    \item A novel CAM abstraction in MLIR that 
    supports different CAMs types and search operations (see Section~\ref{subsec:camdialect}).  
    \item Transformation passes to optimize for latency, power, and array utilization (see Section~\ref{subsec:camdialect}).
    \item A comprehensive evaluation of the generated code, including validation and comparison to a GPU target and the hand-crafted implementations (see Section~\ref{sec:eval}).
\end{itemize}

Our evaluation of C4CAM demonstrates that the generated code achieves comparable results to hand-crafted designs. 
We also showcase the capabilities of C4CAM in performing design space exploration on different CAM architectures.

\section{Background and related work}
\label{sec:background}
This section presents background on the MLIR framework and CAM-based structures and describes our proposed architecture. It also motivates the need for automatic compilation tools by explaining the challenges in the state-of-the-art programming models for CAMs. 

\subsection{MLIR compiler infrastructure}
\label{subsec:MLIR}
MLIR is a framework that enables representing and transforming intermediate representations (IR) at various abstraction levels, catering to diverse application domains and heterogeneous hardware targets~\cite{mlir}.  It offers a customizable IR, with minimal built-in features, enabling compiler developers to incorporate their own abstractions. This empowers them to optimize for specific domains or targets by leveraging matching techniques at the appropriate levels of abstraction. 

MLIR consists of a collection of reusable abstractions organized into \emph{dialects}. Each dialect incorporates custom types, operations, and attributes, which serve as fundamental building blocks of the IR. In MLIR, values are associated with compile-time known types, while attributes provide compile-time information linked to operations. Dialects in MLIR maintain preconditions for transformation validity within their IR, reducing the complexity and cost of analysis passes. Dialects are typically designed for specific domains (e.g., \texttt{linalg} for linear algebra, \texttt{TOSA} for tensor operations), representations (e.g., \texttt{affine} for the polyhedral model, \texttt{scf} for control flow), or targets (e.g., \texttt{gpu}, \texttt{cim}). The \texttt{llvm} dialect models LLVM IR constructs. Abstractions in MLIR can be progressively lowered (from high-level domain-specific dialects to low-level platform-specific dialects) and raised~\cite{prograising}. 

\subsection{Content addressable memories}
\label{subsec:cam}
\begin{figure}[tb]
\centering
\includegraphics[scale=0.45]{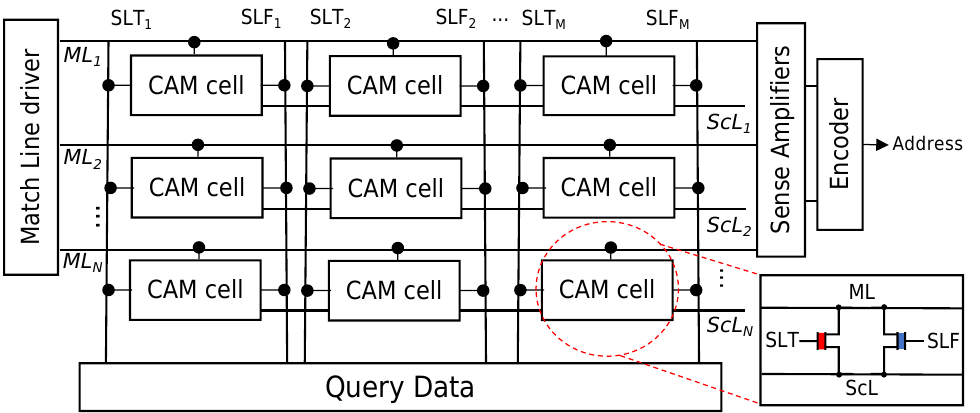}
\caption{Structure of a FeFET-based CAM array \cite{yin2020fecam}}
\label{fig_cam}
\vspace{-0.7cm}
\end{figure}

Content addressable memories (CAMs) are one type of CIM fabric that enables fast and energy-efficient search operations. 
CAMs support two main functions: search, which identifies the memory entries that match the input query, and write, which stores data entries in the memory cells. 
With CAMs, parallel searches can be performed on all stored data in memory in constant time (O(1)). 
The most common type of CAM is the ternary CAM (TCAM), where each element of queries and stored data can be either 0, 1, or don't care (\lq{x}\rq), which is a wildcard state matching both 0 and 1. 
Figure~\ref{fig_cam} illustrates a TCAM array with $R$ rows and $C$ columns. Each cell in a row is connected to a common match line (ML) and stores one of the tree states. During a search operation, each cell $C_{ij}$ in row $i$ performs an XNOR operation between its content and the query element $q_{j}$. The ML implements a logic OR operation of all the cells in the row to determine the result for that row. 

Different sensing circuits can be designed to realize different match schemes, such as EX, BE, and TH. EX search is the fastest search type due to its simple sensing requirement, whereas best match search reports the row with the least number of mismatching cells and is widely used for nearest neighbor search. To find the best match, more sophisticated sensing circuits are needed, e.g., analog-digital-converters or a winner-take-all circuit, with the latter being more energy and area efficient but limited to finding the best matches only within a certain number of mismatch cells~\cite{imani2019searchd}. 

CAM's efficient data retrieval capabilities, i.e., outputting the addresses of stored data that match the input search query based on a specific distance metric, make it highly suitable for applications that rely heavily on large-scale matching or search operations. Recent works have demonstrated the use of CAMs in various fields, e.g., bioinformatics~\cite{Barkam2023hdgim}, high dimensional computing~\cite{kazemi2022achieving}, reinforcement learning~\cite{li2022associative}, few-shot learning~\cite{laguna2019ferroelectric} and recommender systems~\cite{li2022imars}.

\subsection{Accelerator architecture} 
\label{subsec:arch}
\begin{figure}
    \centering
    \includegraphics[width=\linewidth]{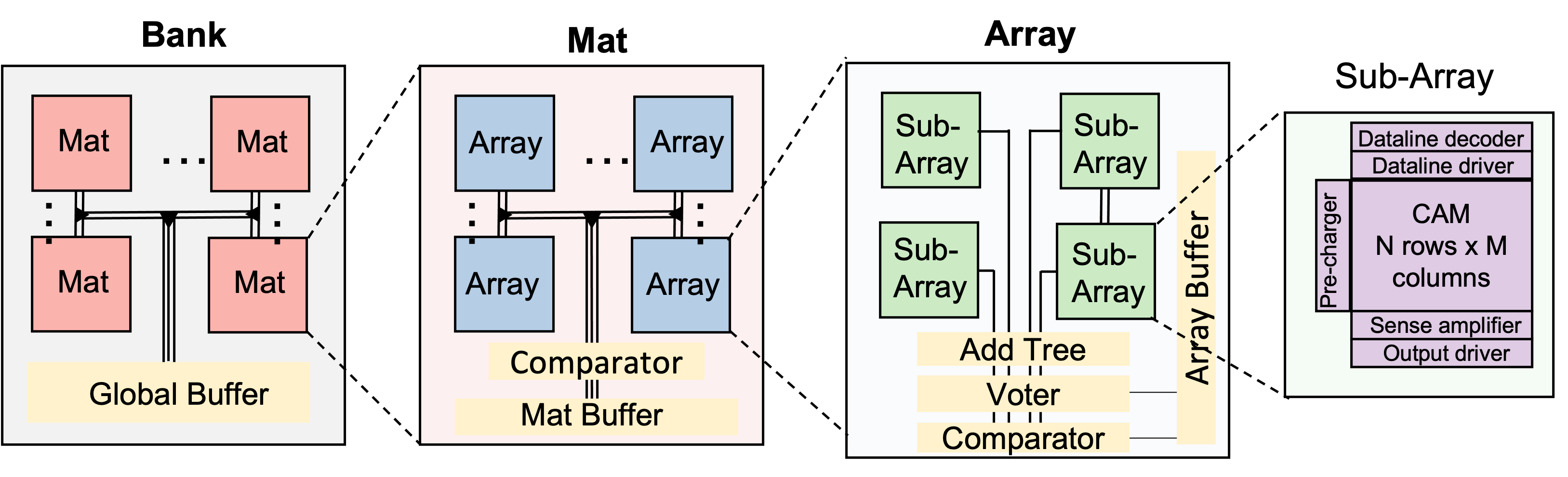}
    \caption{Hierarchical structure of a CAM-based accelerator}
    \label{fig:CAMarch}
\vspace{-0.6cm}
\end{figure}

For this work, we consider a general CAM-accelerator design based on the state-of-the-art~\cite{kazemi2022achieving}. As illustrated in Figure~\ref{fig:CAMarch}, the CAM structure is organized into a four-level hierarchy comprising $B$ banks, each bank containing $T$ mats where each mat consists of $A$ CAM arrays which are further partitioned into $S$ subarrays. The subarrays can be operated and accessed independently. 
This hierarchical organization allows for scalable and flexible computation, as the number of banks, mats, and arrays can be allocated according to the computational requirements of each application.
Within each bank, all mats and arrays can perform parallel search operations using the $S$ CAM subarrays either in a sequential or parallel manner, providing further granularity for parallel processing and resource allocation.
$B$ banks operate independently to allow for task-level parallelism. RecSys~\cite{li2022imars}, for instance, can profit from CAMs in both filtering and ranking stages, where each stage executes different tasks on different banks in parallel. 

\subsection{State-of-the-art programming models for CIM-CAMs}
\label{subsec:motiv}
Most of the current proposals for compilers targeting CIM architectures primarily focus on crossbar-based accelerators \cite{siemieniuk2021occ, khan2023cinm} or general-purpose logic-in-memory architectures \cite{soeken2016mig}.
While the fundamental programming model of the CIM abstraction introduced by CINM~\cite{khan2023cinm} is generic and can be applied to any CIM architecture, it does not provide support for CAM accelerators. CINM primarily focuses on facilitating arithmetic and logic operations, whereas CAMs are specifically designed to handle search operations that involve computing distances, similarities, or comparisons. 
Proposals for supporting CAM-based accelerators are relatively rare, such as DT2CAM \cite{rakka2023dt2cam}, a framework for mapping and simulating decision trees onto TCAMs. However, this mapping tool does not generalize to other comparison-intensive kernels and requires programmers with a deep understanding of both the application and the accelerator architecture. 
Therefore, there is a considerable demand for a generalized framework capable of efficiently handling lowering high-level language programs for diverse input applications. This framework should also incorporate CAM array optimizations, such as selective row precharging, to generate optimized code for the underlying architecture. In the following section, we introduce how hierarchical C4CAM framework effectively addresses this gap.

\section{The C4CAM framework}
\label{sec:camcompiler}
This section presents C4CAM, including the abstractions, and the lowering, analysis and optimization passes.  

\begin{figure*}[tbh]
\centering
\includegraphics[scale=0.65]{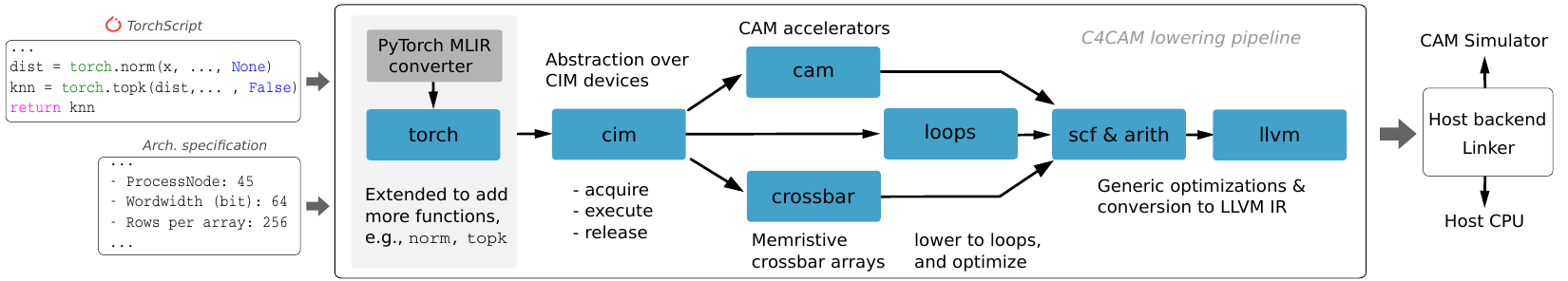}
\caption{A high level overview of the C4CAM  flow}
\label{fig:compflow}
\vspace{-0.6cm}
\end{figure*}

\subsection{An overview of the compilation flow}
\label{subsec:overview}
Figure \ref{fig:compflow} shows a high-level overview of C4CAM.
The TorchScript functions chosen for offloading to the CAM accelerators are transformed into MLIR's representation using the PyTorch MLIR converter (see Section~\ref{subsec:c4cam-frontEnd}). This produces the \texttt{Torch} IR which is the entry point into C4CAM and encompasses ATen tensor library operations. 

Torch MLIR is then lowered to the \texttt{cim} abstraction which is a comprehensive solution for various CIM technologies, taking over the shared responsibilities of host-device interfacing and device mapping (see Section~\ref{subsec:cimdialect}). The \texttt{cim} abstraction has been previously investigated in~\cite{occ} and~\cite{khan2023cinm}, where a programming model for CIM devices was introduced. C4CAM extends this abstraction by incorporating the necessary analysis for CAM devices.

To enable the mapping, \texttt{cim} supports partitioning, rewriting, or modifying the functions to include device-compatible  sizes and operations, which the low-level dialects can then process. Subsequently, the \texttt{cim} dialect is either lowered to \texttt{cam} or \texttt{loop}, or other device dialects.

The \texttt{cam} dialect and other device dialects at the same level, such as \texttt{crossbar}, offer an abstraction for programming and executing functions on the target device (see Section~\ref{subsec:camdialect}). The \texttt{cam} dialect also provides transformation passes that enable mapping and optimization of the selected kernel, while accounting for the concrete  
 hierarchy and other characteristics of CAM-based architectures.

\subsection{Architecture specification}
\label{subsec:c4cam-archSpec}
In addition to the input application, C4CAM takes the architectural configuration as input, as shown in Figure~\ref{fig:compflow}. This configuration outlines the hierarchy of the proposed architecture (as discussed in Section~\ref{subsec:arch}), as well as the access mode for each level of the hierarchy, whether it supports sequential or parallel accesses. Note that all active rows within a subarray are accessed in parallel. However, through selective row accessing~\cite{zukowski1997use}, it is possible to activate and pre-charge only a  subset of rows within a subarray. Furthermore, this input file specifies the optimization target which can be set to latency, power, or array utilization.

\subsection{C4CAM front end}
\label{subsec:c4cam-frontEnd}
The PyTorch MLIR converter~\cite{torch_mlir} is responsible for converting Python code written in TorchScript. However, certain operations from the ATen library, particularly those used in search-based applications such as \texttt{norm} and \texttt{topk}, are not supported. From the CAMs perspective, these are essential primitives in any input application. Since C4CAM is built upon the MLIR framework and this is the only available front end that enables lowering TorchScript input to the MLIR torch dialect, we extend the frontend to support the \texttt{norm} and \texttt{topk} primitives that are commonly accelerated on CAM arrays.

 \begin{figure}[h]
         \begin{subfigure}{\columnwidth}
                \centering
\begin{lstlisting}[language=Python,basicstyle=\tiny\ttfamily,linerange={90-94},breaklines=true,postbreak=\mbox{\textcolor{red}{$\hookrightarrow$}\space},frame=tb,linewidth=\columnwidth]
    def forward(self, input: Tensor, dot: bool = False) -> Tensor:
        others = self.weight.transpose(-2, -1)
        matmul = torch.matmul(input, (others))
        values, indices = torch.ops.aten.topk(matmul, 1, largest=False)
        return indices 
\end{lstlisting}
                \vspace{-0.3cm}
                \caption{PyTorchScript code for HDC dot similarity}
                \label{fig:torchScript}
        \end{subfigure}
        \hfill
        \begin{subfigure}{\columnwidth}
                \centering
\begin{lstlisting}[language=mlir,basicstyle=\tiny\ttfamily,linerange={9-11},breaklines=true,postbreak=\mbox{\textcolor{red}{$\hookrightarrow$}\space},frame=tb,linewidth=\columnwidth]
    %1 = torch.aten.transpose.int %0, %int-2, %int-1 : !torch.vtensor<[10,8192],f32>, !torch.int, !torch.int -> !torch.vtensor<[8192,10],f32>
    %2 = torch.aten.mm %arg0, %1 : !torch.vtensor<[10,8192],f32>, !torch.vtensor<[8192,10],f32> -> !torch.vtensor<[10,10],f32>
    %values, %indices = torch.aten.topk %2, %int1, %int-1, %false, %true : !torch.vtensor<[10,10],f32>, !torch.int, !torch.int, !torch.bool, !torch.bool -> !torch.vtensor<[10,1],f32>, !torch.vtensor<[10,1],f32>
\end{lstlisting}
                 \vspace{-0.3cm}
                \caption{\texttt{Torch} IR for HDC dot similarity.}
                \label{fig:torchIR}
        \end{subfigure}
        \vspace{-0.6cm}

\caption{Python and MLIR representations of HDC similarity}
\label{fig:py2MLIR}
 \vspace{-0.6cm}
\end{figure}

\subsection{C4CAM progressive lowering}
\label{subsec:c4cam-flow}
The compilation flow begins with the \texttt{Torch} dialect, as depicted in Figure~\ref{fig:compflow}. This dialect includes most of the ATen tensor library operations. 
To enable support for the \texttt{cim} abstraction, we have introduced a \texttt{torch-to-cim} conversion pass. This pass lowers the operations that are compatible with the \texttt{cim} abstraction. Examples of these operations include \texttt{topk}, \texttt{norm}, \texttt{sub}, and \texttt{matmul}, which can be executed individually or as part of a kernel on a CIM device.

To demonstrate how the IR of the application transforms at each hierarchy level, we use the similarity kernel in hyperdimensional computing (HDC) as a running example. Figure~\ref{fig:torchScript} shows the TorchScript code of the input kernel. Figure~\ref{fig:torchIR} presents its MLIR representation at the \texttt{Torch} abstraction as produced by the MLIR PyTorch front end. 
The conversion from \texttt{Torch} to \texttt{cim} is accomplished with target-agnoistic transformations. The outcome of the conversion primarily showcases the interface with a generic CIM device.

\subsubsection{The extended \texttt{cim} abstraction}
\label{subsec:cimdialect}
\texttt{cim} is the primary high-level dialect encompassing device-supported operations and essential transformations required to prepare kernels to run on target devices.
This abstraction is mainly responsible for: (i) analyzing the input code to identify CIM-amenable primitives that can be offloaded to the accelerator, (ii) if a CIM-executable pattern is identified but the operand sizes exceed the array sizes specified in the given architecture, dividing the input into smaller partitions to ensure compatibility with the accelerator, and (iii) providing an abstract programming model to enable the execution of kernels on a device.  

\begin{figure}[tbh]
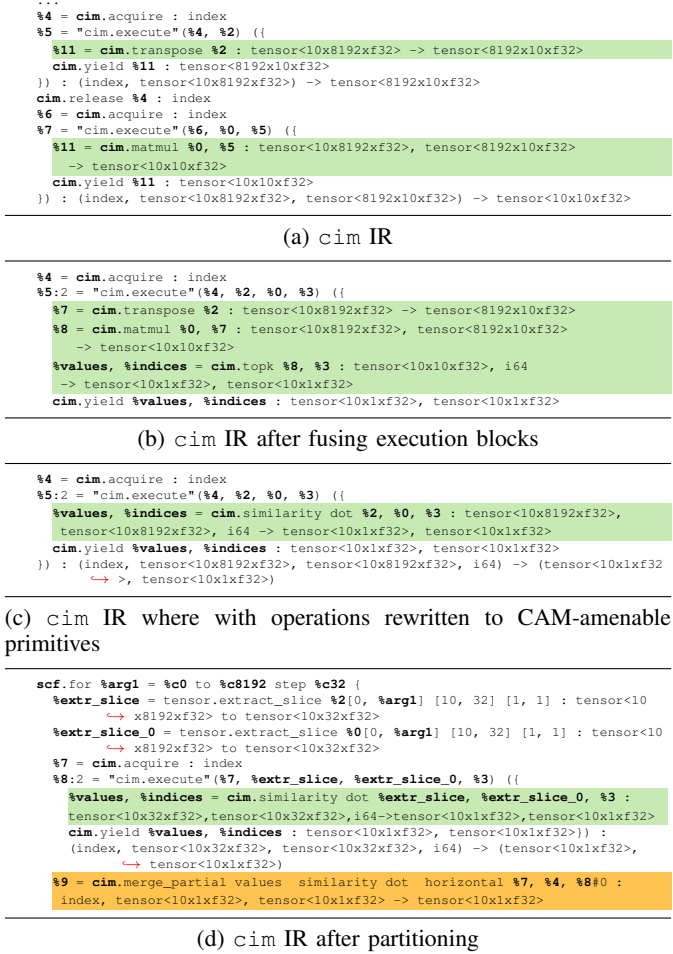

        \begin{subfigure}{\columnwidth}
                \centering
                
\begin{lstlisting}[escapechar=\$,language=mlir,basicstyle=\tiny\ttfamily,linerange={12-24},breaklines=true,postbreak=\mbox{\textcolor{red}{$\hookrightarrow$}\space},frame=tb,linewidth=\columnwidth]
    ...
    %4 = cim.acquire : index
    %5 = "cim.execute"(%4, %2) ({
      $\hlhlhl$%11 = cim.transpose %2 : tensor<10x8192xf32> -> tensor<8192x10xf32>
      cim.yield %11 : tensor<8192x10xf32>
    }) : (index, tensor<10x8192xf32>) -> tensor<8192x10xf32>
    cim.release %4 : index
    %6 = cim.acquire : index
    %7 = "cim.execute"(%6, %0, %5) ({
      $\hlhlhl$%11 = cim.matmul %0, %5 : tensor<10x8192xf32>, tensor<8192x10xf32> 
      $\hlhlhl$  -> tensor<10x10xf32>
      cim.yield %11 : tensor<10x10xf32>
    }) : (index, tensor<10x8192xf32>, tensor<8192x10xf32>) -> tensor<10x10xf32>
\end{lstlisting}
                 \vspace{-0.3cm}

                \caption{\texttt{cim} IR}
                \label{fig:cimIR}
        \end{subfigure}
        \hfill
        \begin{subfigure}{\columnwidth}
                \centering
\begin{lstlisting}[escapechar=\$,language=mlir,basicstyle=\tiny\ttfamily,linerange={12-19},breaklines=true,postbreak=\mbox{\textcolor{red}{$\hookrightarrow$}\space},frame=tb,linewidth=\columnwidth]
    %4 = cim.acquire : index
    %5:2 = "cim.execute"(%4, %2, %0, %3) ({
      $\hlhlhl$%7 = cim.transpose %2 : tensor<10x8192xf32> -> tensor<8192x10xf32>
      $\hlhlhl$%8 = cim.matmul %0, %7 : tensor<10x8192xf32>, tensor<8192x10xf32>
      $\hlhlhl$   -> tensor<10x10xf32>
      $\hlhlhl$%values, %indices = cim.topk %8, %3 : tensor<10x10xf32>, i64 
      $\hlhlhl$ -> tensor<10x1xf32>, tensor<10x1xf32>
      cim.yield %values, %indices : tensor<10x1xf32>, tensor<10x1xf32>
\end{lstlisting}
                 \vspace{-0.3cm}

                \caption{\texttt{cim} IR after fusing execution blocks}
                \label{fig:cimfused}
        \end{subfigure}
        \hfill
         \begin{subfigure}{\columnwidth}
                \centering
\begin{lstlisting}[escapechar=\$,language=mlir,basicstyle=\tiny\ttfamily,linerange={12-17},breaklines=true,postbreak=\mbox{\textcolor{red}{$\hookrightarrow$}\space},frame=tb,linewidth=\columnwidth]
    %4 = cim.acquire : index
    %5:2 = "cim.execute"(%4, %2, %0, %3) ({
      $\hlhlhl$%values, %indices = cim.similarity dot %2, %0, %3 : tensor<10x8192xf32>,
      $\hlhlhl$ tensor<10x8192xf32>, i64 -> tensor<10x1xf32>, tensor<10x1xf32>
      cim.yield %values, %indices : tensor<10x1xf32>, tensor<10x1xf32>
    }) : (index, tensor<10x8192xf32>, tensor<10x8192xf32>, i64) -> (tensor<10x1xf32>, tensor<10x1xf32>)
\end{lstlisting}
                 \vspace{-0.3cm}

                \caption{\texttt{cim} IR where with operations rewritten to CAM-amenable primitives}
                \label{fig:cimIR_rewrite}
        \end{subfigure}
        \hfill
        \begin{subfigure}{\columnwidth}
                \centering
\begin{lstlisting}[escapechar=\$,language=mlir,basicstyle=\tiny\ttfamily,linerange={17-27},breaklines=true,postbreak=\mbox{\textcolor{red}{$\hookrightarrow$}\space},frame=tb,linewidth=\columnwidth]
    scf.for %arg1 = %c0 to %c8192 step %c32 {
      %extr_slice = tensor.extract_slice %2[0, %arg1] [10, 32] [1, 1] : tensor<10x8192xf32> to tensor<10x32xf32>
      %extr_slice_0 = tensor.extract_slice %0[0, %arg1] [10, 32] [1, 1] : tensor<10x8192xf32> to tensor<10x32xf32>
      %7 = cim.acquire : index
      %8:2 = "cim.execute"(%7, %extr_slice, %extr_slice_0, %3) ({
        $\hlhlhlmedium$%values, %indices = cim.similarity dot %extr_slice, %extr_slice_0, %3 : 
        $\hlhlhlmedium$tensor<10x32xf32>,tensor<10x32xf32>,i64->tensor<10x1xf32>,tensor<10x1xf32>
        cim.yield %values, %indices : tensor<10x1xf32>, tensor<10x1xf32>}) : 
        (index, tensor<10x32xf32>, tensor<10x32xf32>, i64) -> (tensor<10x1xf32>, tensor<10x1xf32>)
      $\glglgl$%9 = cim.merge_partial values  similarity dot  horizontal %7, %4, %8#0 : 
      $\glglgl$ index, tensor<10x1xf32>, tensor<10x1xf32> -> tensor<10x1xf32>
\end{lstlisting}
                 \vspace{-0.3cm}

                \caption{\texttt{cim} IR after partitioning}
                \label{fig:cimpartitioned}
        \end{subfigure}
        \caption{\texttt{cim} IR of the HDC similarity function, after different analysis and optimization passes}
        \label{fig:cimdiffIR}
        \vspace{-0.6cm}
        \end{figure}

The programming model employed in C4CAM's \texttt{cim} abstraction for CIM devices is derived from the concept proposed in \cite{khan2023cinm}. This model encompasses three main functions: 
To allocate an accelerator, \texttt{cim} uses the \texttt{cim.acquire} function that returns a handle to the device. 
The \texttt{cim.execute} function uses this handle and specifies the operations that are to be executed on this accelerator. Finally, the device is released using the \texttt{cim.release} function. 
For CAM architectures, we show how these functions are lowered to different CAM functions in Section~\ref{subsec:camdialect}. 
Figure~\ref{fig:cimIR} shows the IR for the running example at the \texttt{cim} abstraction, which can be produced by running the conversion pass \texttt{torch-to-cim} at the \texttt{Torch} abstraction. As the \texttt{Torch} abstraction does not, and is not supposed to, specify the kernel type, the fundamental assumption of the \texttt{torch-to-cim} conversion is that each supported operation can be executed on a separate (non-)CIM device.
Since all the torch operations are supported by the \texttt{cim} dialect (as they are part of the dot similarity), they are lowered to their corresponding \texttt{cim} versions.

\textbf{Pattern matching and fusing:} \texttt{cim} implements analysis and optimization passes to recover patterns that can be offloaded to a CIM accelerator and, when possible, optimizes them for the target. The analysis pass identifies blocks containing operations that cannot be directly lowered to the accelerator and fuses them.
Once the code analysis is complete, the execution blocks can be transformed and offloaded to CIM accelerators, or they can follow the standard MLIR pipeline to generate \texttt{llvm} code for execution on the host processor.

 \vspace{-0.3cm}
    \begin{algorithm}[htb]

    \scriptsize
        \SetKwData{Result}{\texttt{\%res}}
        \SetKwData{ResultPrime}{\(\Result^\prime\)}
        \SetKwData{Operands}{\texttt{\%op:N}}
        \SetKwData{OperandsPrime}{\(\Operands^\prime\)}
        \SetKwData{Part}{\texttt{\%part}}
        \SetKwData{InRects}{in\_rects} \SetKwData{OutRect}{out\_rect}
        \SetKwFunction{Uninitialized}{Uninitialized}
        \SetKwFunction{ExtractSlice}{ExtractSlice}
        \SetKwFunction{CloneOps}{CloneOps}
        \SetKwFunction{InsertSlice}{InsertSlice}

        \KwIn{cim::ExecuteOp op}
        
        \emph{DotProdSimPattern} = \{(transpose()$\rightarrow$(v1)), (matmul(v1)$\rightarrow$(v2)), (topk(v2)$\rightarrow$(v3))\}\;
        \emph{EuclNormPattern} = \{(sub()$\rightarrow$(v1)), (norm(v1)$\rightarrow$(v2)), (topk(v2)$\rightarrow$(v3))\}\;
        \emph{CosSimPattern} = \{(norm()$\rightarrow$(v1)), (norm()$\rightarrow$(v2)), (transpose()$\rightarrow$(v3)), (matmul(v3)$\rightarrow$(v4)), (div(v4, v2, v1)$\rightarrow$(v5))\}\;

        opList = op.getBody().getOperations()\;
        opSize = opList.size()\;
        \If{opSize == 4 }{
            \Return {similarDFG(opList, DotProdSimPattern) || similarDFG(opList, EuclNormPattern)}\;
        } \ElseIf {opSize == 6} {
            \Return {similarDFG(opList, CosSimPattern)}\;
        }
        \Return Failure()\;
        
        \caption{\small SimilarityMatching Function.}
        \label{algo:similiarity_matching}
    \end{algorithm}

 \vspace{-0.3cm}

Algorithm~\ref{algo:similiarity_matching} demonstrates how the \texttt{cim} dialect examines execution blocks to verify whether they correspond to a similarity search operation.
The algorithm checks if the number of operations and the data flow of the operations align with predefined supported patterns. 
For instance, the patterns for similarity based on a dot product, Euclidean norm, and the cosine function are defined in Lines 1, 2, and 3, respectively. 
The \texttt{cim-fuse-ops} pass, when enabled with the similarity flag indicating the search for similarity operations, employs this algorithm to identify code blocks that match the criteria and subsequently replace their operations with the \texttt{cim.similarity} operation.

Figure~\ref{fig:cimIR} shows the base \texttt{cim} IR produced by the \texttt{torch-to-cim} conversion pass, while Figure~\ref{fig:cimIR_rewrite} showcases the result obtained after applying the fusion pass to Figure~\ref{fig:cimfused}.

\textbf{Compulsory partitioning:} 

Kernels often require more space than what the processing elements (PE) of the target can support. To overcome this limitation, the kernel is partitioned according to the size supported by a PE. For instance, in a CAM system, the smallest block within the system is the subarray. Therefore, when partitioning the application, it is important to consider this level of granularity and divide it accordingly.
To support this, C4CAM includes a partitioning transformation within the \texttt{cim} abstraction. This transformation can be likened to tiling in compiler terminology, with some hardware-specific considerations. 
It enables the efficient partitioning of kernels to facilitate their execution on the device(s), but requires an abstraction to accumulate partial results.
To this end, the \texttt{cim} dialect includes the \texttt{cim.merge\_partial} operation. 
It considers both the type of operation for which partial results are generated and the direction in which these results are accumulated. The partitioned version of Figure~\ref{fig:cimIR_rewrite} for a device of size 32x32 is demonstrated in Figure~\ref{fig:cimpartitioned}.

The \texttt{cim} abstraction focuses on identifying operations that can be offloaded to the CAM accelerator and partitions them based on the subarray size. It does not address the mapping of an input application or its partitions onto the CAM accelerator, nor does it incorporate any device-specific optimizations. The latter are performed by the device-specific \texttt{cam} abstraction, which is discussed in the next section. 

\subsubsection{The \texttt{cam} abstraction}
\label{subsec:camdialect}
To convert the \texttt{cim} IR into the \texttt{cam} IR, C4CAM introduces the \texttt{cim-to-cam} conversion pass. This pass requires the specification of the target CAM device type (e.g., ACAM, TCAM or MCAM) as an input parameter, which also determines the search type and metric to be utilized during the conversion process. 
The \texttt{cam} dialect is responsible for mapping the high-level functions from the \texttt{cim} dialect to the CAM-device calls. After applying this conversion pass, occurrences of a sequence of \texttt{cim.acquire}, \texttt{cim.execute}, and \texttt{cim.release} working on the same device handle are substituted with calls to allocate a simple system consisting of a bank, a mat, an array, and a single subarray. This system executes the targeted execution block intended to run on the chosen CAM.

 More concretely, the \texttt{cam.alloc\_bank} function is used to allocate a CAM bank, taking the row and column sizes of the desired CAM size as parameters. Furthermore, allocating a mat from the bank and a CAM array from the mat, and a subarray from an array is accomplished using the \texttt{cam.alloc\_mat}, \texttt{cam.alloc\_array}, and \texttt{cam.alloc\_subarray} functions, respectively.
Similarly, the \texttt{cim.execute} function is lowered into three CAM function calls: \texttt{cam.write\_value}, \texttt{cam.search}, and \texttt{cam.read\_value}.
The write operation programs the CAM arrays with the input data. The search operation performs the actual search on the data based on the specified search type and metric. The supported search types include exact match, best match, and range match, while the available distance metrics are Euclidean and Hamming.
The read operation reads the values and indices of the search results from the device.

  \begin{figure}[tbh]
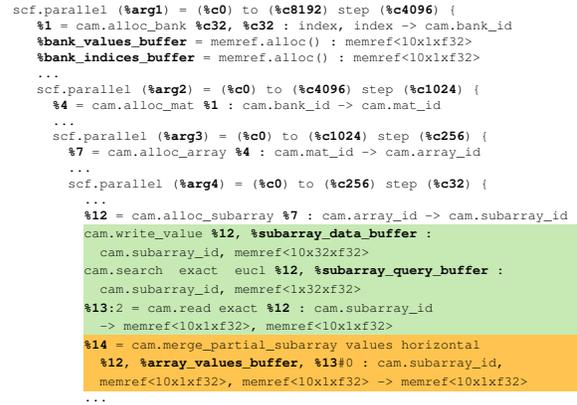

                \vspace{-0.3cm}
                \centering
\begin{lstlisting}[escapechar=\$,language=mlir,basicstyle=\tiny\ttfamily,linerange={2-25},breaklines=true,postbreak=\mbox{\textcolor{red}{$\hookrightarrow$}\space},frame=tb,linewidth=\columnwidth]
   scf.parallel (%arg1) = (%c0) to (%c8192) step (%c4096) {
      %1 = cam.alloc_bank %c32, %c32 : index, index -> cam.bank_id
      %bank_values_buffer = memref.alloc() : memref<10x1xf32>
      %bank_indices_buffer = memref.alloc() : memref<10x1xf32>
      ...
      scf.parallel (%arg2) = (%c0) to (%c4096) step (%c1024) {
        %4 = cam.alloc_mat %1 : cam.bank_id -> cam.mat_id
        ...
        scf.parallel (%arg3) = (%c0) to (%c1024) step (%c256) {
          %7 = cam.alloc_array %4 : cam.mat_id -> cam.array_id
          ...
          scf.parallel (%arg4) = (%c0) to (%c256) step (%c32) {
            ...
            %12 = cam.alloc_subarray %7 : cam.array_id -> cam.subarray_id
            $\hlhlhlsmall$cam.write_value %12, %subarray_data_buffer : 
            $\hlhlhlsmall$  cam.subarray_id, memref<10x32xf32>
            $\hlhlhlsmall$cam.search  exact  eucl %12, %subarray_query_buffer : 
            $\hlhlhlsmall$  cam.subarray_id, memref<1x32xf32>
            $\hlhlhlsmall$%13:2 = cam.read exact %12 : cam.subarray_id 
            $\hlhlhlsmall$  -> memref<10x1xf32>, memref<10x1xf32>
            $\glglglsmall$%14 = cam.merge_partial_subarray values horizontal 
            $\glglglsmall$  %12, %array_values_buffer, %13#0 : cam.subarray_id, 
            $\glglglsmall$  memref<10x1xf32>, memref<10x1xf32> -> memref<10x1xf32>
            ... 
\end{lstlisting}
                \vspace{-0.3cm}
                \caption{\texttt{cam} IR after mapping}
                \label{fig:campartitioned}
        
\label{fig:camIR}
\vspace{-0.3cm}

\end{figure}

The original program underwent partitioning at the CIM dialect without considering the hierarchy.
This approach was chosen because dealing with synchronization and accumulation of partial results across different levels of the hierarchy often requires hardware-specific information, which goes against the principles of the \texttt{cim} dialect.
To map an application onto the CAM abstraction, the \texttt{cam-map} pass within the \texttt{cam} dialect can be employed. This pass transforms the application into a nested loop structure according to the provided specifications, incorporating the required hardware calls at each loop level.

The code in Figure~\ref{fig:campartitioned} shows the mapping 
The required devices (bank, tile, array, subarray) and buffers for storing the results (values and indices) are allocated at each level of the loop. Additionally, the necessary functions responsible for accumulating these partial results are called.
In cases where the system size precisely matches the data size, the levels of the nested loop, starting from the outermost level, iterate over the banks, mats within each bank, arrays within each bank, and subarrays within each array.
However, if the data size exceeds the system's capacity, an additional loop is introduced. This loop includes iteration over banks, allowing the system to be called multiple times to process the data effectively. 

The \texttt{cim} to \texttt{cam} conversion pass also performs bufferization of tensors involved in executing a kernel on the CAM. This process determines how the memory is handled between the host and the device. During the process of lowering from \texttt{cam} to \texttt{scf} and subsequently to \texttt{llvm}, the \texttt{cam} operations are mapped to function calls of a CAM simulator.

\textbf{Built-in optimizations:}
C4CAM provides an extensible and flexible framework that enables future research in code optimizations and auto-tuning. 
Currently, the framework uses simple heuristics to optimize for different metrics, namely, for latency/performance, power consumption and device utilization.  
This is enabled by device-specific transformations that can be further composed by performance engineers. 
 For example, in order to minimize latency, C4CAM prioritizes maximizing the utilization of parallel-executing arrays in the system. In contrast, to minimize power consumption C4CAM reduces  the number of enabled subarrays at a time inside an array.
For devices that support selective search and if the standard data placement is not feasible due to a smaller number of rows compared to the array, multiple batches of data can be placed on the same array. By utilizing selective search, different queries can be searched on corresponding rows of the same array in multiple cycles. 
As demonstrated in Section~\ref{subsec:dse_fixed}, employing the same hierarchy specification (mat, array, and subarray sizes) consistently leads to longer latency. However, the impact on latency varies depending on the dimensions of the subarray, which subsequently alters the number of banks. This variation can either increase or decrease the energy consumption.

\section{Evaluation}
\label{sec:eval}
This section presents our experimental setup 
and gives a detailed analysis of the code generated with C4CAM.

\vspace{-0.2cm}
\subsection{Experimental setup}
\label{subsec:setup}

\subsubsection{System setup and technology parameters}
\label{sss:tech_params}

For the CAM technology parameters, we consider the 2FeFET CAM design proposed in \cite{yin2020fecam} at the 45nm technology node. Energy and latency numbers for TCAM and MCAM operations were extracted from Eva-CAM \cite{liu2022eva}. Since we are varying the array size for design space exploration, the search latency can vary from $860ps$ to $7.5ns$ for array sizes of $16\times16$ and $256\times256$, respectively.
For the GPU results, we use the NVIDIA Quadro RTX 6000 GPU (16 nm process). The power consumption is measured using the NVIDIA System Management Interface \texttt{nvidia-smi}, and energy is derived thereof.

\subsubsection{Simulation infrastructure}
\label{sss:functional_sim}
We use the same simulation infrastructure as in~\cite{li2022imars}~\cite{kazemi2022achieving}. 
It models the architecture and performs functional simulation of the functions called by C4CAM. 
We extend the simulator to include performance and energy estimation.
To handle large data dimensions and entry sizes, the extended simulator allows for fine-grain control of the hierarchy, and models CAM queries to obtain energy and latency based on real hardware behavior. The functional simulation generates CAM outputs that can be analyzed to assess application accuracy. Moreover, the tool supports different underlying CAM designs and performs architectural modeling for chip-level estimations. Additionally, it estimates the performance cost of additional peripherals to provide application-level results. 
For a fair comparison, we were also granted access to the same simulation and evaluation parameters.

\subsubsection{Benchmarks}
\label{sss:benches}
We evaluate C4CAM on two benchmarks.
The first one is \textit{K-nearest neighbor (KNN)}, a popular algorithm used for classification, regression, and anomaly detection tasks. It works by identifying the $K$ closest training examples in the feature space to a given test sample. It is especially interesting because of its versatility and interopretability, with no training required. 
KNNs are both memory and computationally expensive, making their scalability and performance strongly limited on conventional systems.  We evaluated KNN on chest X-Ray images from the Pneumonia dataset.

The second benchmark is \textit{hyperdimensional computing (HDC)} which is a framework inspired by the human brain's ability to process information. It utilizes high-dimensional vectors known as hypervectors as a fundamental building block. Hypervectors are large binary vectors with thousands of dimensions. We evaluated HDC on the MNIST dataset with 8k dimensions. 
HDC on MNIST is a commonly used benchmark that allows us to validate the results of C4CAM and compare them against existing manual implementations on GPUs and CAM-based accelerators.
The PyTorch GPU implementation operates on \texttt{int32} elements.

\subsection{Validation}
\label{subsec:validation}
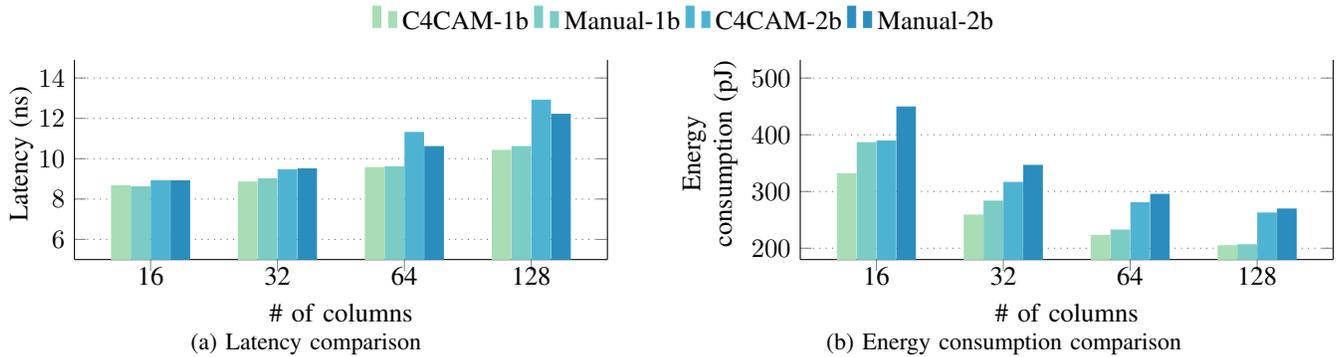
\begin{figure*}[hbt!]

        \begin{subfigure}{\columnwidth}
        \vspace{-0.3cm}

        \centering
        \scalebox{1}{
        \pgfplotsset{compat = newest}
\pgfplotsset{major grid style={dotted,aluminium2!50!black}}
\begin{tikzpicture}[baseline={(current axis.south)}]
\begin{axis}
[
    width=0.8\columnwidth,
    height=\figheight,
    scale only axis,
    ybar=0.5pt, 
    enlargelimits=0.2,
    enlarge y limits={upper, value=0.1},
    ylabel style={align=center},
    y label style={at={(-0.06,0.53)}},
    ylabel= Latency (ns),
    legend style={draw=none, fill=none},
    bar width=7pt,
    legend columns=4,
    ymin=5, ymax=14,
    height=0.3\textwidth,
    ymajorgrids=true,
    grid style=dashed,
    axis x line*=bottom,
    x tick label style={xshift=.0em, yshift=-.2em, rotate=0,anchor=center},
    yminorticks=true,
    legend style={at={(1.75, 1.3)},anchor=north east},
    xlabel={\# of columns},
    symbolic x coords={16, 32, 64, 128},
    xtick=data,
]
\addplot+ [blind_safe_three_scheme_seven_colors_grnblu] coordinates {
(16, 8.65)
(32, 8.84)
(64, 9.56)
(128, 10.41)
}; 
\addplot+ [blind_safe_four_scheme_seven_colors_grnblu] coordinates {
(16, 8.6)
(32, 9)
(64, 9.6)
(128, 10.6)
}; 

\addplot+ [blind_safe_five_scheme_seven_colors_grnblu] coordinates {
(16, 8.9)
(32, 9.45)
(64, 11.3)
(128, 12.9)
}; 

\addplot+ [blind_safe_six_scheme_seven_colors_grnblu] coordinates {
(16,  8.9)
(32, 9.5)
(64, 10.6)
(128, 12.2)
}; 
\legend{C4CAM-1b, Manual-1b, C4CAM-2b, Manual-2b}
\end{axis}
\end{tikzpicture}
        }
        \vspace{-0.2cm}

        \caption{Latency comparison}
        \label{fig:validation_lat}
        \end{subfigure}
        \hfill
        \begin{subfigure}{\columnwidth}
        \vspace{-0.3cm}

                \centering
                \pgfplotsset{compat = newest}
\pgfplotsset{major grid style={dotted,aluminium2!50!black}}
\begin{tikzpicture}[baseline={(current axis.south)}]
\begin{axis}
[
    width=0.8\columnwidth,
    height=\figheight,
    scale only axis,
    ybar=0.5pt, 
    enlargelimits=0.2,
    enlarge y limits={upper, value=0.1},
    ylabel style={align=center},
    y label style={at={(-0.1,0.53)}},
    ylabel= Energy \\consumption (pJ),
    legend style={draw=none, fill=none},
    bar width=7pt,
    legend columns=4,
    ymin=180, ymax=500,
    height=0.3\textwidth,
    ymajorgrids=true,
    grid style=dashed,
    axis x line*=bottom,
    x tick label style={xshift=.0em, yshift=-.2em, rotate=0,anchor=center},
    yminorticks=true,
    legend style={at={(1, 1.2)},anchor=north east},
    xlabel={\# of columns},
    symbolic x coords={16, 32, 64, 128},
    xtick=data,
]
\addplot+ [blind_safe_three_scheme_seven_colors_grnblu] coordinates {
(16, 331.4)
(32, 258.4)
(64, 222.5)
(128, 204.3)
};
\addplot+ [blind_safe_four_scheme_seven_colors_grnblu] coordinates {
(16, 386)
(32, 283)
(64, 232)
(128, 206)
}; 
\addplot+ [blind_safe_five_scheme_seven_colors_grnblu] coordinates {
(16, 389)
(32, 316)
(64, 280)
(128, 262)
}; 
\addplot+ [blind_safe_six_scheme_seven_colors_grnblu] coordinates {
(16, 449)
(32, 346)
(64, 295)
(128, 269)
}; 
\legend{}
\end{axis}
\end{tikzpicture}
                \vspace{-0.2cm}

                 \caption{Energy consumption comparison}
                 
                \label{fig:validation_energy}
        \end{subfigure}
\caption{C4CAM validation against manual designs~\cite{kazemi2022achieving}.}
\label{fig:validation}
\vspace{-0.6cm}

\end{figure*}

In order to validate the C4CAM framework, we use the CAM-design and hand-optimized mapping from~\cite{kazemi2022achieving} as a baseline. 
We generate code for binary and multi-bit implementations of HDC for different CAM architectures, i.e., with array sizes of $32 \times C$ where $C$ is varied to 16, 32, 64, and 128. 
For this evaluation, we use the same system configuration as in the baseline, i.e., four mats per bank, four arrays per mat, eight sub-arrays per array, and as many banks as needed to store the whole dataset.

The validation results for latency and energy are shown in Figure~\ref{fig:validation_lat} and Figure~\ref{fig:validation_energy}, respectively.
In this experiment, the observed deviation in the latency and the energy consumption is, on average (geomean), 0.9\% and 5.5\%, respectively (notice that the y-axes do not start at 0 for better visualization).
These small deviations can be attributed to slight differences in the versions of the simulation environment rather than to fundamental differences in the implementations. 
Hence, C4CAM effectively matches the quality of the manual designers. 

To understand the latency results in Figure~\ref{fig:validation_lat}, it is important to note that all search operations happen in parallel, and the ML discharges more slowly for larger columns. 
As for the energy numbers shown in Figure~\ref{fig:validation_energy}, larger $C$ leads to lower energy consumption because fewer peripherals and fewer levels (arrays, mats, and banks) are required as $C$ increases. 
Moreover, as observed in~\cite{kazemi2022achieving}, we corroborate that binary implementations are more energy efficient than multi-bit ones. This improvement is associated with the higher ML and data line voltages of the multi-bit implementations.

\noindent\textbf{GPU comparison:} We compared end-to-end performance against the GPU setup using the CIM setup described in~\cite{kazemi2022achieving}. The improvement in execution time is $48\times$ which deviates only by 5\% compared to the manual design, while the energy consumption improvement translates to $46.8\times$ which is nearly the same since CAMs contribute minimally to the overall energy consumption in their CIM system.

\begin{table}[htb]
\scriptsize
\centering
\caption{Number of subarrays used to implement HDC.}
\label{tab:CP_RWR}
\begin{tabular}{c|c|c|c|c|c}
\label{table:resource_utilization}
 & $16\times16$ & $32\times32$ & $64\times64$ & $128\times128$ & $256\times256$ \\ \hline
cam-based & 512 & 256 & 128 & 64 & 32  \\ \hline
cam-density & 512 & 86 & 22 & 6 & 2 \\ \hline
\end{tabular}
\vspace{-0.35cm}
\end{table}

\vspace{-0.2cm}
\subsection{Design space exploration}
\label{subsec:dse}
\subsubsection{Fixed architectural parameters}
\label{subsec:dse_fixed}
\pgfplotsset{compat = newest}
\pgfplotsset{major grid style={dotted,aluminium2!50!black}}
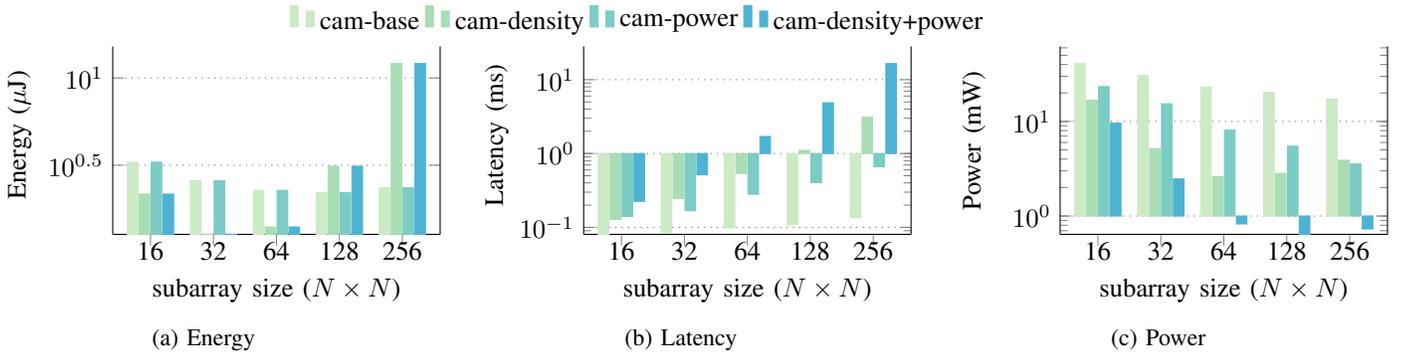
\begin{figure*}
\begin{subfigure}{0.3\textwidth}
\begin{tikzpicture}
\begin{axis}
[
    width=0.8\textwidth,
    height=0.5\figheight,
    scale only axis,
    ybar=0.5pt, 
    enlargelimits=0.15,
    enlarge y limits={upper, value=0.1},
    ylabel style={align=center},
    y label style={at={(-0.22,0.53)}},
    ylabel= Energy ($\mu$J),
    legend style={draw=none, fill=none},
    bar width=4pt,
    legend columns=4,
    ymode = log,
    log basis y={10},
    ymajorgrids=true,
    grid style=dashed,
    axis x line*=bottom,
    x tick label style={xshift=.0em, yshift=-.2em, rotate=0,anchor=center},
    yminorticks=true,
    legend style={at={(2.7, 1.25)},anchor=north east},
    xlabel={subarray size ($N\times N$)},
    symbolic x coords={16, 32, 64, 128, 256},
    xtick=data,
]
\addplot+ [blind_safe_two_scheme_seven_colors_grnblu] coordinates {
(16, 3.29E-0)
(32, 2.58E-0)
(64, 2.27E-0)
(128, 2.20E-0)
(256, 2.35E-0)
}; 
\addplot+ [blind_safe_three_scheme_seven_colors_grnblu] coordinates {
(16, 2.16E-0)
(32, 1.27E-0)
(64, 1.40E-0)
(128, 3.12E-0)
(256, 1.21E1)
}; 
\addplot+ [blind_safe_four_scheme_seven_colors_grnblu] coordinates {
(16, 3.29E-0)
(32, 2.58E-0)
(64, 2.27E-0)
(128, 2.20E-0)
(256, 2.35E-0)
}; 
\addplot+ [blind_safe_five_scheme_seven_colors_grnblu] coordinates {
(16, 2.16E-0)
(32, 1.27E-0)
(64, 1.40E-0)
(128, 3.12E-0)
(256, 1.21E1)
}; 
\legend{cam-base, cam-density, cam-power, cam-density+power}
\end{axis}
\end{tikzpicture}
\caption{Energy}
\label{fig:dse:energy}
\end{subfigure}
\hfill
\begin{subfigure}{0.3\textwidth}
\begin{tikzpicture}[baseline={(current axis.south)}]
\begin{axis}
[
    width=0.8\textwidth,
    height=0.5\figheight,
    scale only axis,
    ybar=0.5pt, 
    enlargelimits=0.15,
    enlarge y limits={upper, value=0.1},
    ylabel style={align=center},
    y label style={at={(-0.2,0.53)}},
    ylabel= Latency (ms),
    legend style={draw=none, fill=none},
    bar width=4pt,
    legend columns=4,
    ymode = log,
    log basis y={10},
    ymajorgrids=true,
    grid style=dashed,
    axis x line*=bottom,
    x tick label style={xshift=.0em, yshift=-.2em, rotate=0,anchor=center},
    yminorticks=true,
    legend style={at={(1.8, 1.`)},anchor=north east},
    xlabel={subarray size ($N\times N$)},
    symbolic x coords={16, 32, 64, 128, 256},
    xtick=data,
]
\addplot+ [blind_safe_two_scheme_seven_colors_grnblu] coordinates {
(16, 8.04E-02)
(32, 8.40E-02)
(64, 9.85E-02)
(128, 000.108237)
(256, 000.135867)
}; 
\addplot+ [blind_safe_three_scheme_seven_colors_grnblu] coordinates {
(16, 0.128587)
(32, 0.245461)
(64, 0.536939)
(128, 1.104341)
(256, 3.103403)
}; 
\addplot+ [blind_safe_four_scheme_seven_colors_grnblu] coordinates {
(16, 0.140567)
(32, 0.168027)
(64, 0.280537)
(128, 0.402237)
(256, 0.660867)
}; 
\addplot+ [blind_safe_five_scheme_seven_colors_grnblu] coordinates {
(16, 0.224907)
(32, 0.514261)
(64, 1.701739)
(128, 4.867541)
(256, 16.543403)
}; 
\legend{}
\end{axis}
\end{tikzpicture}
\caption{Latency}
\label{fig:dse:latency}
\end{subfigure}
\hfill
\begin{subfigure}{0.3\textwidth}
  \begin{tikzpicture}[baseline={(current axis.south)}]
  \begin{axis}
  [
      width=0.8\textwidth,
    height=0.5\figheight,
    scale only axis,
      ybar=0.5pt, 
      enlargelimits=0.15,
      enlarge y limits={upper, value=0.1},
      ylabel style={align=center},
      y label style={at={(-0.2,0.5)}},
      ylabel= Power (mW),
      legend style={draw=none, fill=none},
      bar width=4pt,
      legend columns=2,
      ymode = log,
      log basis y={10},
      ymajorgrids=true,
      grid style=dashed,
      axis x line*=bottom,
      x tick label style={xshift=.0em, yshift=-.2em, rotate=0,anchor=center},
      yminorticks=true,
      legend style={at={(1,1.2)},anchor=north east},
      xlabel={subarray size ($N\times N$)},
      symbolic x coords={16, 32, 64, 128, 256},
      xtick=data,
  ]
  \addplot+ [blind_safe_two_scheme_seven_colors_grnblu] coordinates {
  (16, 40.96842206)
  (32, 30.75569914)
  (64, 23.06871134)
  (128, 20.32916116)
  (256, 17.2934791)
  }; 
  \addplot+ [blind_safe_three_scheme_seven_colors_grnblu] coordinates {
  (16, 16.79123418)
  (32, 5.170070572)
  (64, 2.615883666)
  (128, 2.82151188)
  (256, 3.890293885)
  }; 

  \addplot+ [blind_safe_four_scheme_seven_colors_grnblu] coordinates {
  (16, 23.42301769)
  (32, 15.38029011)
  (64, 8.10273376)
  (128, 5.470313431)
  (256, 3.55534252)
  };
  
  \addplot+ [blind_safe_five_scheme_seven_colors_grnblu] coordinates {
  (16, 9.60011041)
  (32, 2.467718908)
  (64, 0.825372964)
  (128, 0.64014088)
  (256, 0.729786288)
  };

  \legend{}
  \end{axis}
  \end{tikzpicture}
  \caption{Power}
  \label{fig:dse:power}
  \end{subfigure}
  \caption{Impact of subarray size and C4CAM optimizations on latency, energy, and power.}
\end{figure*}

\begin{table*}[htb]
\vspace{-0.2cm}
\centering
\caption{EDP and power for KNN execution.}
\label{tab:knn}
\begin{tabular}{c|c|c|c|c|c||c|c|c|c|c}
\multicolumn{1}{c}{}& \multicolumn{5}{c||}{\textbf{EDP (nJ$\cdot$s)}} & \multicolumn{5}{c}{\textbf{POWER (W)}} \\ \hline
subarray size & 16x16 & 32x32 & 64x64 & 128x128 & 256x256 & 16x16 & 32x32 & 64x64 & 128x128 & 256x256 \\ \hline
cam-based & 0.75 & 0.30 & 0.15 & 0.08 & 0.05 & 44.14 & 16.30 & 5.97 & 2.34 &  0.86 \\ \hline
cam-power &  1.32 & 0.61 & 0.44 & 0.29 & 0.23 & 25.23 & 8.15 & 2.10 & 0.66 & 0.19 \\ \hline
\end{tabular}
\vspace{-0.4cm}
\end{table*}

While we could reproduce the results of single manual designs, the automation provided by C4CAM allows for quick exploration of different software and hardware implementations.
To demonstrate this, we evaluated systems consisting of sub-arrays with sizes of $R \times C$, where $C = R$ assuming values of 16, 32, 64, 128, and 256 with different configurations for the same, as outlined in Section~\ref{sss:tech_params}, namely:
\begin{itemize}
  \item \textit{cam-base}: In this configuration, applications are allocated to the CAM accelerator without incorporating the optimizations discussed in Section~\ref{subsec:camdialect}. In this setup, parallel execution is enabled at each level.  
  \item \textit{cam-power}: In this configuration, we implement a restriction on the maximum number of sub-arrays activated concurrently. Specifically, for each application, we have chosen to enable only one sub-array per array at a time.

  \item \textit{cam-density}: This configuration demonstrates the impact of employing selective search~\cite{zukowski1997use} to enhance both the utilization of arrays and the system's overall capacity, as shown in Table~\ref{table:resource_utilization}.
    \item \textit{cam-power+density}: This configuration imposes limitations on the number of enabled sub-arrays at a time. Simultaneously, it incorporates selective search technique to enhance the system's capacity.
\end{itemize}
For all sub-array sizes, the configuration remains consistent, with 4 mats per bank, 4 arrays per mat, and 8 sub-arrays per array. We use as many banks as needed to accommodate the input data. 
Figure~\ref{fig:dse:energy} and Figure~\ref{fig:dse:latency} illustrate the energy consumption and latency of the configurations mentioned above respectively when executing the HDC application on the MNIST dataset with 8k dimensions.

In the \emph{cam-power} configuration, only one sub-array within the array is active at a time. With a sub-array of size $16\times16$, the power consumption is reduced to approximately $0.57\times$ with respect to the base configuration (Figure~\ref{fig:dse:power}).
Similarly, the power requirement for the largest array size is merely 20\% of the base configuration. However, this reduction in power consumption results in increased latency. For instance, executing the application on a $32\times32$-sized subarray incurs a latency increase of approximately $2\times$ compared to the baseline. As the array size increases, the latency rises, reaching up to $4.86\times$ the baseline for the largest sub-array size. The overall energy consumption remains the same between the two configurations, \emph{cam-power} and \emph{cam-base}.

The analysis of the KNN benchmark is similar to the analysis for HDC.  For space reasons we summarize the results in Table~\ref{tab:knn} for EDP and power. 
The absolute values of energy and latency are considerably higher than in the HDC case. 
This is simply due to the sheer size of the Pneumonia dataset, requiring many banks in the CAM accelerator.

The \emph{cam-density} configuration uses selective search to improve resource utilization, as shown in Table~\ref{table:resource_utilization}. In the case of the smallest array size ($16\times16$), the execution time is less than twice compared to the base configuration. This trend scales further, and with the largest subarray size ($256\times256$), the execution time is nearly $23\times$ longer compared to the \emph{cam-base} configuration.
The energy consumption for subarray sizes ranging from $16\times16$ to $64\times64$ in the cam-density configuration is, on average, $0.6\times$ that of the corresponding sub-array size in the baseline configuration. However, for sub-arrays of $128\times128$ or $256\times256$, the energy consumption increases compared to the baseline, reaching $1.4\times$ and $5.1\times$, respectively. It is worth noting that by fixing the system configuration and enabling selective search, the number of banks required for application execution is reduced, thus reducing the overall power consumption.

The \emph{cam-power-density} configuration combines the approaches of both \emph{cam-power} and \emph{cam-density} to achieve the most significant reduction in power consumption. A $16\times16$-sized subarray utilizes only 23.4\% of the base power, while the largest sub-array requires only 4.2\% of the base power. However, this reduction in power consumption comes at the cost of significantly increasing the execution time. In the case of the largest subarray configuration, the execution time is approximately $121\times$ longer compared to the base configuration.

\subsubsection{Iso-capacity analysis}
With the iso-capacity experiments, we investigate the relationship between energy consumption and latency by changing the size of subarrays and the number of subarrays per array while keeping the capacity fixed to $2^{16}$ TCAM cells per array.
To achieve this, we modify the subarray size, starting from $256\times256$ which corresponds to one subarray per array, and gradually decrease it to $16\times16$, resulting in 256 subarrays per array.
The numbers of arrays per mat and mats per bank are fixed as in the previous sections.
It is important to note that these systems are not iso-area since each subarray has its own set of peripherals. 
This means that as the size of the subarrays is reduced, more peripherals are needed, and chip area increases.

\begin{figure}[t]
\centering
  \begin{subfigure}[b]{0.6\textwidth}
  \begin{tikzpicture}
\begin{axis}
[
    width=0.7\columnwidth,
    height=.8\figheight,
    ybar=0.5pt, 
    enlargelimits=0.08,
    enlarge y limits={upper, value=0.1},
    ylabel style={align=center},
    y label style={at={(-0.15,0.5)}},
    ylabel= Latency (ms),
    legend style={draw=none, fill=none},
    bar width=4.5pt,
    legend columns=4,
    ymode = log,
    log basis y={10},
    ymajorgrids=true,
    grid style=dashed,
    axis x line*=bottom,
    x tick label style={xshift=.0em, yshift=-.2em, rotate=0,anchor=east},
    yminorticks=true,
    legend style={at={(1,1.25)},anchor=north east},
    xlabel={subarray size ($N\times N$)},
    symbolic x coords={16, 32, 64, 128, 256},
    xtick=data,
]
\addplot+ [blind_safe_three_scheme_seven_colors_grnblu] coordinates {
(16, 5.84E-02)
(32, 6.94E-02)
(64, 9.12E-02)
(128, 0.115557)
(256, 0.150507)
}; 
\addplot+ [blind_safe_four_scheme_seven_colors_grnblu] coordinates {
(16, 8.17E-02)
(32, 0.175189)
(64, 0.490091)
(128, 1.198037)
(256, 3.290795)
}; 
\addplot+ [blind_safe_five_scheme_seven_colors_grnblu] coordinates {
(16, 3.590539)
(32, 2.594389)
(64, 2.986091)
(128, 2.810837)
(256, 3.290795)
}; 
\legend{iso-base, iso-density, iso-density+power}
\end{axis}
\end{tikzpicture}
\vspace{-0.2cm}
\caption{Latency}

\label{fig:isocap:latency}
\end{subfigure}
\begin{subfigure}[b]{0.6\textwidth}

\begin{tikzpicture}
\begin{axis}
[
    width=0.7\columnwidth,
    height=.8\figheight,
    ybar=0.5pt, 
    enlargelimits=0.08,
    enlarge y limits={upper, value=0.1},
    ylabel style={align=center},
    y label style={at={(-0.15,0.4)}},
    ylabel= Power (mW),
    legend style={draw=none, fill=none},
    bar width=4.5pt,
    legend columns=7,
    ymode = log,
    log basis y={10},
    ymajorgrids=true,
    grid style=dashed,
    axis x line*=bottom,
    x tick label style={xshift=.0em, yshift=-.0em, rotate=45,anchor=east},
    yminorticks=true,
    legend style={at={(1,1.1)},anchor=north east},
    xlabel={subarray size ($N\times N$)},
    symbolic x coords={16, 32, 64, 128, 256},
    xtick=data,
]
\addplot+ [blind_safe_three_scheme_seven_colors_grnblu] coordinates {
(16, 42.09468782)
(32, 31.30287625)
(64, 23.57260442)
(128, 20.12701186)
(256, 18.5582772)
}; 
\addplot+ [blind_safe_four_scheme_seven_colors_grnblu] coordinates {
(16, 19.80106486)
(32, 5.850260107)
(64, 2.7236024)
(128, 2.744546458)
(256, 3.758513951)
}; 

\addplot+ [blind_safe_five_scheme_seven_colors_grnblu] coordinates {
(16, 0.450771539)
(32, 0.395046015)
(64, 0.4470099236)
(128, 1.169782784)
(256, 3.758513951)
}; 

\legend{}
\end{axis}
\end{tikzpicture}
\vspace{-0.2cm}
\caption{Power}

\label{fig:isocap:energy}
\end{subfigure}
  \caption{Impact of optimizations on iso-capacity setups.}
\vspace{-0.7cm}
\end{figure}
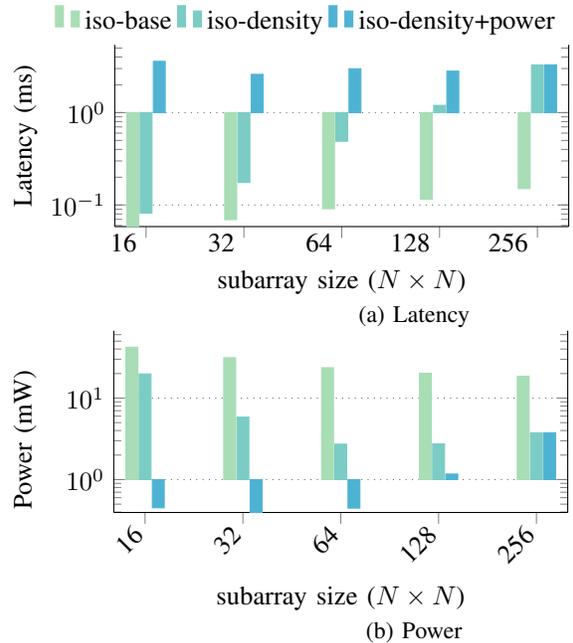

Figure~\ref{fig:isocap:energy} shows that the energy consumption in \textit{iso-base} remains nearly constant across  subarray sizes.
Moreover, \textit{cam-density} and \textit{cam-power+density}, on average, achieve $1.75\times$ energy improvement over \textit{iso-capacity-base}, except for large subarray sizes like $128\times128$ and $256\times25$. 
The total execution time across different subarray sizes also varies within a moderate range, i.e., from $58 \mu s$ for $16\times16$ to $150 \mu s$ for $256\times256$ , as shown in Figure~\ref{fig:isocap:energy}. 
Again, as the search latency increases for larger columns, the execution time also increases despite the consistent number of cells within an array.
As for the \textit{cam-density} and \textit{cam-power+density} transformations, 
Figure~\ref{fig:isocap:energy} shows a significant decrease in power consumption, offering potential CAM configuration that can be used in power-constrained system setups.

\section{Conclusions}
\label{sec:conclusions}
We present C4CAM, a framework that enables the exploration of CAM configurations and seamless code generation from high-level TorchScript code.
We introduce an MLIR abstraction named \texttt{cam} that is specifically tailored for CAM-based accelerators. This abstraction provides control knobs that allow for the tuning of various metrics by adjusting the mapping of applications to the CAM arrays. 
To validate the effectiveness of C4CAM, we compare our results with those obtained from a hand-crafted design and demonstrate that C4CAM produces comparable results. 
Moreover, we demonstrate C4CAM capabilities by automatically generating implementations optimized for performance, power and device utilization. 
Finally, we show how C4CAM retargetability facilitates design space exploration by varying architectural parameters without any application recoding effort. 
The architecture specification supported by C4CAM, along with its compilation flow, also enables the specification of heterogeneous systems. However, determining the optimal mapping strategy for heterogeneous systems based on different optimization targets remains a subject for future research.

\section*{Acknowledgments}
This work is partially funded by the German Research Council (DFG) through the HetCIM(502388442) projects. 

\bibliographystyle{IEEEtran}
\bibliography{main.bib}

\end{document}